\documentclass[galaxies,article,accept,moreauthors,pdftex,10pt,a4paper]{Definitions/mdpi} 

\def\la{\ifmmode\stackrel{<}{_{\sim}}\else$\stackrel{<}{_{\sim}}$\fi} 
\def\ga{\ifmmode\stackrel{>}{_{\sim}}\else$\stackrel{>}{_{\sim}}$\fi}

\firstpage{1} 
\pubvolume{xx}
\issuenum{1}
\articlenumber{5}
\pubyear{2020}
\copyrightyear{2020}
\history{Received: date; Accepted: date; Published: date}

\usepackage{amssymb}
\usepackage{Definitions/aas_macros}

\Title{Magnetism Science with the Square Kilometre Array}

\Author{George Heald~$^{1,}$*~\orcidA{}, Sui Ann Mao~$^{2}$, Valentina Vacca~$^{3}$~\orcidE{}, Takuya Akahori~$^{4}$~\orcidC{}, Ancor Damas-Segovia~$^{5}$~\orcidG{}, B.~M.~Gaensler~$^{6}$~\orcidF{}, Matthias Hoeft~$^{7}$, Ivan Agudo~$^{5}$~\orcidJ{}, Aritra Basu~$^{8}$~\orcidO{}, Rainer Beck~$^{2}$, Mark Birkinshaw~$^{9}$~\orcidM{}, Annalisa Bonafede~$^{10,11}$, Tyler L. Bourke~$^{12}$~\orcidP{}, Andrea Bracco~$^{13}$~\orcidR{}, Ettore Carretti~$^{11}$, Luigina Feretti~$^{11}$~\orcidK{}, J.~M. Girart$^{14,15}$~\orcidXE{}, Federica Govoni~$^{3}$~\orcidH{}, James A. Green~$^{1}$~\orcidXB{}, JinLin Han~$^{16,17,18}$~\orcidI{}, Marijke Haverkorn~$^{19}$~\orcidL{}, Cathy Horellou~$^{20}$~\orcidN{}, Melanie Johnston-Hollitt~$^{21}$~\orcidXD{}, Roland Kothes~$^{22}$, Tom Landecker$^{22}$, B\l{}a\.zej Nikiel-Wroczy\'nski~$^{23}$~\orcidB{}, Shane P. O'Sullivan~$^{24}$~\orcidX{}, Marco Padovani~$^{25}$~\orcidQ{}, Fr\'ed\'erick Poidevin$^{26,27}$~\orcidD{}, Luke Pratley~$^{6}$, Marco Regis~$^{28,29}$~\orcidU{}, Christopher John Riseley~$^{10,11}$~\orcidXC{}, Tim Robishaw~$^{22}$~\orcidXF{}, Lawrence Rudnick~$^{30}$~\orcidV{}, Charlotte Sobey~$^{1}$~\orcidXA{}, Jeroen M. Stil~$^{31}$~\orcidT{}, Xiaohui Sun~$^{32}$~\orcidS{}, Sharanya Sur~$^{33}$, A. Russ Taylor~$^{34,35,36}$, Alec Thomson~$^{1}$~\orcidZ{}, Cameron L. Van Eck~$^{6}$~\orcidY{}, Franco Vazza~$^{10,11}$, Jennifer L. West~$^{6}$~\orcidW{}, and the SKA Magnetism Science Working Group~$^{37}$}

\AuthorNames{George Heald, Sui Ann Mao, Valentina Vacca, Takuya Akahori, Ancor Damas-Segovia, B.~M.~Gaensler, Matthias Hoeft, Ivan Agudo, Aritra Basu, Rainer Beck, Mark Birkinshaw, Annalisa Bonafede, Tyler L. Bourke, Andrea Bracco, Ettore Carretti, Luigina Feretti, J.~M. Girart, Federica Govoni, James A. Green, JinLin Han, Marijke Haverkorn, Cathy Horellou, Melanie Johnston-Hollitt, Roland Kothes, Tom Landecker, B\l{}a\.zej Nikiel-Wroczy\'nski, Shane P. O'Sullivan, Marco Padovani, Fr\'ed\'erick Poidevin, Luke Pratley, Marco Regis, Christopher John Riseley, Tim Robishaw, Lawrence Rudnick, Charlotte Sobey, Jeroen M. Stil, Xiaohui Sun, Sharanya Sur, A. Russ Taylor, Alec Thomson, Cameron L. Van Eck, Franco Vazza, Jennifer L. West, and the SKA Magnetism Science Working Group}

\address{%
$^{1}$ \quad CSIRO Astronomy and Space Science, PO Box 1130, Bentley WA 6102, Australia\\
$^{2}$ \quad Max Planck Institute for Radio Astronomy, Auf dem H\"ugel 69, 53121, Bonn, Germany\\
$^{3}$ \quad INAF - Osservatorio Astronomico di Cagliari, Via della Scienza 5, 09047 Selargius (CA), Italy\\
$^{4}$ \quad Mizusawa VLBI Observatory, National Astronomical Observatory of Japan (NAOJ), 2-21-1 Osawa, Mitaka, Tokyo 181-8588, Japan\\
$^{5}$ \quad Instituto de Astrof\'isica de Andaluc\'ia (CSIC), Glorieta de la Astronom\'ia, 18008 Granada, Spain \\
$^{6}$ \quad Dunlap Institute for Astronomy and Astrophysics, University of Toronto, 50 St George Street, Toronto, ON M5S 3H4, Canada\\
$^{7}$ \quad Th\"uringer Landessternwarte, Sternwarte 5, 07778 Tautenburg, Germany\\
$^{8}$ \quad Fakult\"at f\"ur Physik, Universit\"at Bielefeld, Postfach 100131, 33501 Bielefeld, Germany\\
$^{9}$ \quad HH Wills Physics Laboratory, University of Bristol, Tyndall Avenue, Bristol BS8 1TL, UK\\
$^{10}$ \quad Dipartimento di Fisica e Astronomia, Universit\`a degli Studi di Bologna, via P. Gobetti 93/2, 40129 Bologna, Italy\\
$^{11}$ \quad INAF - Istituto di Radioastronomia, Via Gobetti 101, 40129 Bologna, Italy\\
$^{12}$ \quad SKA Organization, Jodrell Bank, Lower Withington, Macclesfield SK11 9FT, UK\\
$^{13}$ \quad Rudjer Bo\v{s}kovi\'c Institute, Bijeni\v{c}ka cesta 54, 10000, Zagreb, Croatia\\
$^{14}$ \quad Institut de Ciencies de l'Espai (CSIC), 08193 Cerdanyola del Vall\`es, Catalonia, Spain\\
$^{15}$ \quad Institut d'Estudis Espacials de Catalunya (IEEC), 08034 Barcelona, Catalonia\\
$^{16}$ \quad National Astronomical Observatories, Chinese Academy of Sciences, Jia-20 DaTun Road, ChaoYang District, Beijing 100101, China\\
$^{17}$ \quad School of Astronomy, University of Chinese Academy of Sciences, Beijing 100049, China\\
$^{18}$ \quad CAS Key Laboratory of FAST, NAOC, Chinese Academy of Sciences, Beijing 100101, China\\
$^{19}$ \quad Department of Astrophysics/IMAPP, Radboud University Nijmegen, P.O. Box 9010, 6500 GL Nijmegen, Netherlands\\
$^{20}$ \quad Chalmers University of Technology, Dept of Space, Earth and Environment, Onsala Space Observatory, 439 92 Onsala, Sweden\\
$^{21}$ \quad International Centre for Radio Astronomy Research, Curtin University, Bentley, WA 6102, Australia\\
$^{22}$ \quad National Research Council Canada, Herzberg Programs in Astronomy \&\ Astrophysics, Dominion Radio Astrophysical Observatory, P.O. Box 248, Penticton, BC V2A 6J9, Canada\\
$^{23}$ \quad Astronomical Observatory of the Jagiellonian University, ul. Orla 171, 30-244 Krak\'ow, Poland\\
$^{24}$ \quad Centre for Astrophysics and Relativity, Dublin City University, Glasnevin, Ireland\\
$^{25}$ \quad INAF - Osservatorio Astrofisico di Arcetri, Largo E. Fermi 5, 50125 Firenze, Italy\\
$^{26}$ \quad Instituto de Astrofis\'ica de Canarias, 38200 La Laguna, Tenerife, Spain\\
$^{27}$ \quad Departamento de Astrof\'isica, Universidad de La Laguna (ULL), 38206 La Laguna, Tenerife, Spain\\
$^{28}$ \quad Dipartimento di Fisica, Universit\`a degli Studi di Torino, Via P. Giuria 1, 10125 Torino, Italy\\
$^{29}$ \quad INFN - Istituto Nazionale di Fisica Nucleare, Sezione di Torino, Via P. Giuria 1, 10125 Torino, Italy\\
$^{30}$ \quad Minnesota Institute for Astrophysics, School of Physics and Astronomy, University of Minnesota, 116 Church Street SE, Minneapolis, MN 55455, USA\\
$^{31}$ \quad Department of Physics and Astronomy, The University of Calgary, 2500 University Drive NW, Calgary, AB T2N 1N4, Canada\\
$^{32}$ \quad Department of Astronomy, Yunnan University, and Key Laboratory of Astroparticle Physics of Yunnan Province, Kunming 650091, China\\
$^{33}$ \quad Indian Institute of Astrophysics, 2nd Block, Koramangala, Bangalore 560034, India\\
$^{34}$ \quad Department of Astronomy, University of Cape Town, Private Bag X3, Rondebosch 7701, South Africa\\
$^{35}$ \quad Department of Physics and Astronomy, University of the Western Cape, Private Bag X17, Bellville 7535, South Africa\\
$^{36}$ \quad Inter-University Institute for Data Intensive Astronomy, Private Bag X3, Rondebosch 7701, South Africa\\
$^{37}$ \quad \url{https://astronomers.skatelescope.org/science-working-groups/magnetism/}
}

\corres{Correspondence: george.heald@csiro.au; Tel.: +61 (0)8 6436 8758}

\abstract{The Square Kilometre Array (SKA) will answer fundamental questions about the origin, evolution, properties, and influence of magnetic fields throughout the Universe. Magnetic fields can illuminate and influence phenomena as diverse as star formation, galactic dynamics, fast radio bursts, active galactic nuclei, large-scale structure, and Dark Matter annihilation. Preparations for the SKA are swiftly continuing worldwide, and the community is making tremendous observational progress in the field of cosmic magnetism using data from a powerful international suite of SKA pathfinder and precursor telescopes. In this contribution, we revisit community plans for magnetism research using the SKA, in the light of these recent rapid developments. We focus in particular on the impact that new radio telescope instrumentation is generating, thus advancing our understanding of key SKA magnetism science areas, as well as the new techniques that are required for processing and interpreting the data. We discuss these recent developments in the context of the ultimate scientific goals for the SKA era.}


\begin{document}

\section{Introduction}

Cosmic magnetism has traditionally been a relatively specialised field, but is increasingly recognised as a domain where new progress is crucial to gain greater understanding of broader astrophysical phenomena such as the star formation process, galaxy evolution, the physics of phenomena related to active galactic nuclei, galaxy clusters and large-scale structure, and the evolution of the early Universe. Following steady progress over the last several decades through a diverse range of observational tracers and theoretical approaches, a leap forward in the radio domain is anticipated with the development of the Square Kilometre Array (SKA). The first phase of the SKA will comprise two interferometric radio telescope arrays: a low-frequency array in Western Australia (SKA1-LOW) observing from $50-350\,\mathrm{MHz}$ with 131,072 broadband log-periodic dipoles organised in 512 stations separated by a maximum baseline of 65~km; and a mid-frequency array in South Africa (SKA1-MID) observing at least from $350-1760$ and $4600-15300\,\mathrm{MHz}$ with 197 offset-Gregorian antennas separated by a maximum baseline of 150~km. A full description of the SKA ``Design Baseline'' is provided on the project website\footnote{\url{https://astronomers.skatelescope.org/documents/}}.

The impact of the SKA on magnetism science was first broadly considered in a volume dedicated to the SKA science case about 16 years ago \cite{gaensler_etal_2004,feretti_etal_2004,feretti_johnston-hollitt_2004,beck_gaensler_2004}. Since then, the SKA Organisation has fostered the development of this science case (and others) within the wide international community through a Cosmic Magnetism Science Working Group (SWG). This SWG was instrumental in developing an updated description of SKA science ambitions, as was comprehensively described about 5 years ago in the two-volume SKA Science Book\footnote{\url{https://pos.sissa.it/cgi-bin/reader/conf.cgi?confid=215}}. An overview of chapters within the field of cosmic magnetism, as well as a description of the headline magnetism ``Rotation Measure (RM) Grid'' survey was provided \cite{johnston-hollitt_etal_2015}. The overview chapter was supplemented with additional detailed contributions in the areas of active galactic nuclei and radio galaxies \cite{agudo_etal_2015,gaensler_etal_2015,laing_2015,peng_etal_2015}, nearby galaxies \cite{beck_etal_2015,heald_etal_2015}, galaxy clusters \cite{bonafede_etal_2015,johnston-hollitt_etal_2015b,govoni_etal_2015}, the Milky Way \cite{haverkorn_etal_2015,Han_etal_2015}, the interstellar medium (ISM) and star formation \cite{dickinson_etal_2015,robishaw_etal_2015}, large-scale structure and the cosmic web \cite{vazza_etal_2015,giovannini_etal_2015}, dark matter \cite{colafrancesco_etal_2015}, and advanced techniques \cite{stil_keller_2015,taylor_etal_2015,vacca_etal_2015}.

Since then, rapid progress in developing our understanding of cosmic magnetism and refining the questions that we will seek to answer with the SKA is already visible based on results that are now starting to flow from SKA pathfinder and precursor activities. This paper seeks to provide an updated snapshot view of the field, on the basis of presentations and discussions at the most recent SKA science meeting\footnote{\url{https://indico.skatelescope.org/event/467/}}, which was held in Manchester (UK) from 8--12 April 2019.

The radio domain\footnote{In this paper, we focus primarily on the frequency range intended to be covered by the SKA: $50\,\mathrm{MHz}\leq\nu\leq24\,\mathrm{GHz}$.} has been very successful for many years in tracing the magnetic fields that pervade the Universe on all physical scales \cite{han_2017,akahori_etal_2018}. Three primary techniques are employed: (i) synchrotron radiation, including its degree of linear and circular polarization; (ii) the Faraday rotation effect induced by intervening magneto-ionic media; and (iii) Zeeman splitting of energy levels in atoms and molecules. A full review of these mechanisms is beyond the scope of this paper; we refer the reader to \cite{heald_2015} for a detailed description of the insights that can be gained from the study of synchrotron emission and Faraday rotation, and \cite{crutcher_kemball_2019,robishaw_etal_2020} for reviews of the power of Zeeman splitting.

This paper is organised as follows. We begin by revisiting the key science areas in the field of cosmic magnetism that we intend to probe with the SKA (\S\,\ref{sec:science}), highlighting advances that have occurred since the publication of the 2015 SKA science book. This progress is now rapidly taking place with an impressive international suite of SKA precursor and pathfinder telescopes (\S\,\ref{sec:pathfinders}), while simultaneously clarifying the capabilities and potential of those instruments. These ongoing developments are not only answering existing scientific questions and prompting new ones, but also serving to clarify the technical considerations that are required to maximise the science achievable using SKA observations (\S\,\ref{sec:tech}), and providing a fresh context for the design of the key surveys anticipated to be delivered by the SKA (\S\,\ref{sec:surveys}).
We conclude the paper with some thoughts about the steps that will be required in the coming years before the SKA enters its operational phase (\S\,\ref{sec:future}).
 
\section{SKA magnetism science cases}\label{sec:science}

Cosmic magnetism has been a key SKA science driver since the early planning phases of the project \citep{carilli_rawlings_2004}, including an early ``RM Grid'' concept (e.g., \cite{gaensler_etal_2004}). The SKA holds great promise because of the central importance of meter- and centimeter-wavelength radio observations for the study of cosmic magnetism. Synchrotron radiation, which traces magnetic fields perpendicular to the line of sight, is typically dominant at these wavelengths. Synchrotron radiation is intrinsically linearly and circularly polarized at a fractional amount that reflects the degree of order of the magnetic field. Faraday rotation and depolarization effects, as well as the conversion of linear into circular polarization (e.g., \cite{Osullivan_etal_2013,gruzinov_levin_2019}), which depend on the magnetic field strength along the line of sight, are observable effects that typically carry detailed behaviour across broad frequency ranges, reflecting the structure of the magnetoionic medium. Furthermore, the Zeeman effect that is observed in molecular lines such as OH, H$_2$O, and, to a lesser extent, CH$_3$OH (as a non-paramagnetic molecule), provides us with an independent measurement of in-situ magnetic fields threading molecular clouds.

The magnetism-oriented chapters\footnote{\url{https://pos.sissa.it/cgi-bin/reader/conf.cgi?confid=215\#session-2111}} of the 2015 SKA Science Book (see \cite{johnston-hollitt_etal_2015} for an overview) describe how modern techniques will be revolutionary for developing a newly detailed observational picture of cosmic magnetism. Specifically, the community now makes use of Faraday Rotation Measure (RM) Synthesis (originally conceived by \cite{burn_1966}, and updated in the context of modern radio telescope technology by \cite{brentjens_debruyn_2005}) and the complementary analysis technique ``Faraday tomography'', through which observed features are associated with structure along the line of sight.
This approach is based on the Faraday depth \cite{burn_1966}
\begin{equation}
    \phi(r)\propto\int_r^0 n_e\,\Vec{B}\cdot \mathrm{d}\Vec{l}
\end{equation}
where $n_e$ is the thermal electron density, $\Vec{B}$ is the magnetic field, $\Vec{l}$ is the line of sight, and the sign convention is that $\phi$ is positive for $\Vec{B}$ directed toward the observer. Although Faraday depth is not equivalent to physical distance, expressing linear polarization as a function of this quantity provides the potential to distinguish multiple magneto-ionic contributions along the line of sight to and through radio sources. However, interpretation of the observed features in a three-dimensional distribution is not a trivial task and requires complementary multi-wavelength observations, simulations and other tools for analysis and interpretation. During the last several years, the power of RM synthesis and Faraday tomography has been demonstrated in various aspects of cosmic magnetism (see for example \cite{haverkorn_etal_2019}), thanks in large part to the wider wavelength coverage of modern radio telescopes. Although Faraday rotation measure and Faraday depth provide combined information about the line-of-sight magnetic field, thermal gas density, and distance, supplementary constraints on the thermal gas density and distance can be provided by multi-wavelength observations in the X-ray and optical bands, respectively, thus isolating and constraining the magnetic field contribution. For impulsive sources such as pulsars and Fast Radio Bursts (FRBs) the dispersion measure ($\mathrm{DM}\propto\int n_e\,\mathrm{d}l$) also provides complementary information about the thermal electron density along the line of sight (e.g., \cite{han_etal_2018}). This multi-wavelength and multi-tracer observational approach has started to take shape through the efforts of various research groups over the past few years.

In this section, we provide brief reviews of the primary magnetism science cases that have been developing, with an emphasis on recent progress that has been made in theoretical and observational studies. The study of cosmic magnetism spans diverse astrophysical domains, and magnetic fields within the corresponding observational targets range over a wide variety of scales, from Mpc down to sub-pc. Overarching these detailed studies are two common questions: 1) what is the origin and evolution of magnetic fields throughout the Universe? and 2) how do magnetic fields illuminate and influence the physical processes in different objects? The SKA will probe magnetoionic media in many different environments including the large-scale structure of the Universe and the intergalactic medium (IGM), large-scale jets and outflows from AGN, the formation and evolution of galaxies and stars, and the properties of the interstellar medium (ISM). 

\subsection{Large-Scale Structure and Cosmology}

\subsubsection{The Cosmic Web}\label{sec:cosmicweb}

The largest structures in the Universe comprise the cosmic web, the network of filaments and sheets that connects galaxy clusters. Constraining the properties of the magnetic field in the cosmic web is at the frontier of cosmic magnetism research. It has been predicted by numerical simulations that the cosmic web is filled with baryons at temperatures ranging from $10^5$ K to $10^7$~K. This warm-hot IGM, or WHIM, is the main candidate for the solution to the so-called `missing baryon problem' \cite{cen_ostriker_1999,dave_etal_2001} in the local Universe, and is cospatial with the intergalactic magnetic field (IGMF). If the IGMF was formed in the early Universe (\S\,\ref{sec:eor}), then its presence can be best traced in voids or on the largest linear scales ($>10\,\mathrm{Mpc}$) in the cosmic web. While in the case of galaxy clusters the presence of significant dynamo amplification of primordial (or galactic) seed fields is beyond doubt, in the IGMF of filaments the level of dynamo amplification and the memory of ancient magnetic seeding events is more debated, and ultimately related to the (unknown) level of plasma turbulence there.

The community is seeking detections of the filamentary magnetised cosmic web through two primary techniques: diffuse, low surface brightness synchrotron emission originating from within the filaments themselves, and through careful analysis of RMs for background sources with lines of sight passing through the large-scale structure.
Recently, a number of radio observations have attempted to constrain the IGMF strength in filaments \cite{brown_etal_2017,vernstrom_etal_2017,vacca_etal_2018,vernstrom_etal_2019,govoni_etal_2019}. RMs from extragalactic polarized sources and Fast Radio Bursts (FRBs; \S\,\ref{sec:frbs}) have been used to place limits on the strength of the IGMF in voids of less than 10~nG \cite{pshirkov_etal_2016,ravi_etal_2016,hackstein_etal_2019,osullivan_etal_2019,osullivan_etal_2020}, which already rules out some of the theoretical predictions \cite{akahori_etal_2014,akahori_etal_2016,vazza_etal_2017}. 
In order to put stringent constraints on the IGMF on the basis of RMs for extragalactic polarized sources, a detailed knowledge of the thermal gas density distribution along the line of sight and its mixing with the magnetic field is required.
Moreover, given the likely low level of RMs from the IGMF in filaments, a careful selection of background sources and excellent sensitivity will be necessary to disentangle the intrinsic contribution from background sources from that of intervening filaments (e.g. \cite{locatelli_etal_2018}), in addition to carefully separating the Galactic RM foreground.
The prospect of constructing RM Grids using FRBs is of particular interest due to the simultaneous availability of rotation and dispersion measures \cite{akahori_etal_2016,vazza_etal_2018}. The SKA will discover hundreds to thousands of localised FRBs \cite{macquart_etal_2015} and therefore deliver RM Grids from FRBs which have the potential to untangle the degeneracy between electron density and magnetic field.

\subsubsection{Early Universe and the Epoch of Reionization}\label{sec:eor}

In the early Universe, primordial magnetogenesis scenarios may require some modifications to the standard theory of inflation and cosmology; such magnetic fields could impact the formation of the large-scale structure in the Universe \citep{kandu_2016}. 
The existence of a pervasive IGMF would introduce non-Gaussian anisotropies in the CMB and could affect the duration of the Epoch of Reionization (EoR) \citep{schleicher_miniati_2011,widrow_etal_2012}, while the EoR itself may be crucial for seeding the IGMF \cite{langer_durrive_2018}. Therefore, observational constraints on the IGMF in the early Universe are an important component for cosmological models. The IGMF in the early Universe may be constrained by the EoR observations that will be undertaken with the SKA \cite{koopmans_etal_2015}, because magnetic fields can alter the expected spatial fluctuations of density and temperature through heating mechanisms from ambipolar diffusion and decay of turbulence in the IGM \cite{minoda_etal_2018}.

\subsubsection{Galaxy Clusters}

Magnetic fields in clusters have historically been revealed through detection of diffuse synchrotron emission from the intracluster medium (ICM), and Faraday RMs of embedded and background radio sources. The number of diffuse radio sources identified in galaxy clusters has increased by a few hundred over the past few years \cite{vanweeren_etal_2019}, largely due to an increase in the prevalence of high-quality observations at low radio frequencies. Radio halos are found to be generally unpolarized, likely due to depolarization effects, limited resolution in Faraday space, and because current interferometers have limited sensitivity and angular resolution. Based on the existence of $\mu\mathrm{G}$-level magnetic fields at cluster centers, intrinsic polarization of radio halos is expected at $15-35\%$ levels \cite{govoni_etal_2013}. Detection of this emission will provide information about the ICM magnetic field power spectrum.

It remains unclear how magnetic fields are distributed and amplified during cluster evolution, and how cosmic rays are (re-)accelerated in this environment. These processes are tightly coupled to the physics of shock waves and magneto-hydrodynamic (MHD) turbulence in the ICM (e.g., \cite{kang_ryu_2013,brunetti_lazarian_2016}). During the formation of galaxy clusters, up to $10^{64}$~ergs are deposited in the ICM on large spatial scales and then on increasingly smaller scales through turbulent cascades.  The strength and coherence length of cluster magnetic fields should depend on the growth stage of MHD turbulence, implying that they vary between merging clusters, regular clusters, and cool-core clusters. 
The mechanism of growth of intracluster magnetic fields is still debated. Initial simulations by \cite{dolag_2000} have shown that its strength saturates during cluster formation, while later work shows that dynamo action can explain the magnetic field strength presently indicated from rotation measure observations \cite{vazza_etal_2018b}.

Cool-core galaxy clusters show higher gas density than merging systems. As the plasma has a very high conductivity, to first order the magnetic field is frozen-in and hence its strength is expected to be higher, implying higher rotation measure values.
This expectation is confirmed by radio observations that indicate magnetic fields with strengths in cluster centres of order $\mu$G and fluctuation scales up to a few hundreds of kpc in merging systems, while observations of relaxed systems indicate magnetic field strengths in cluster centres up to order 10 $\mu$G and fluctuation scales of the order of tens kpc or less (see e.g. \cite{vacca_etal_2018b}). Moreover, magnetic field strength is expected to correlate with gas density and/or temperature \cite{govoni_etal_2017}.

SKA observations of galaxy clusters will be powerful for determining cluster magnetic field properties through RM synthesis if the cluster area is sufficiently sampled by background polarized sources, and especially by diffuse polarized background radio sources to resolve field structures smaller than the separation of sources in the RM Grid (e.g., \cite{loi_etal_2019}). The statistics of the spatial fluctuations of the larger sources can reveal the turbulent properties of the ICM.

\subsubsection{Dark Matter}\label{sec:dm}

Dark matter is a fundamental ingredient of our Universe and of structure formation models, and yet its fundamental nature is elusive to astrophysical probes. Information on the nature and physical properties of the Weakly Interacting Massive Particle (WIMP), the leading candidate for a cosmologically relevant dark matter, can be obtained by studying the astrophysical signals of their annihilation/decay. Among the various electromagnetic signals, electrons and positrons produced by WIMP annihilation generate synchrotron emission in the magnetized atmosphere of galaxy clusters and galaxies, which could be observed as a diffuse radio emission centered on the dark matter halo. A deep search for dark matter radio emission with the SKA in local dwarf galaxies, galaxy regions with low star formation, galaxy clusters (with offset dark matter-baryonic distribution, e.g. the Bullet cluster \cite{markevitch_etal_2004}), and studying angular correlations of the unresolved radio background can be effective in constraining the WIMP mass and annihilation cross-section \cite{colafrancesco_etal_2015}. These limits strongly depend on the magnetic field, which is typically poorly known in the quiescent regions of interest, i.e., in the regions where the astrophysical ``background'' is dim.

Studies with different telescopes have recently attempted the detection of a radio WIMP signal, including the Green Bank Telescope (GBT) \cite{spekkens_etal_2013}, Australia Telescope Compact Array (ATCA) \cite{regis_etal_2014,regis_etal_2017}, LOFAR \cite{vollmann_etal_2019}, and MWA \cite{cook_etal_2020}. They all obtained a null detection, and derived upper limits on the dark matter interaction rate. The SKA will have the capability to determine simultaneously both the magnetic field strength in the dark matter-dominated structures and the dark matter particle properties. It has been shown that the search for WIMP radio signals with the SKA and its precursors (for example in the very relevant case of dwarf spheroidal galaxies \cite{regis_etal_2014}) will progressively close in on the full parameter space of WIMPs, which means it will constrain the dark matter annihilation cross-section below the so-called ``thermal value''.
With the SKA, the uncertainties associated with the description of the ambient medium will dramatically decrease, reducing the astrophysical uncertainties and allowing a more precise determination of the dark matter bounds.

\subsubsection{FRBs}\label{sec:frbs}

The study of Fast Radio Bursts (FRBs) has grown rapidly \cite{lorimer_2018} since the first detection \cite{lorimer_etal_2007} just over a decade ago, with a particular rise in activity associated with localisation efforts \cite{chatterjee_etal_2017,bannister_etal_2019} aimed at resolving outstanding questions about their origin and physical properties. Rotation measures from FRBs can be very useful to help constrain the environment of these enigmatic sources. As a remarkable example, observations of repeating FRB121102 indicated an extremely large RM of $\mathcal{O}(10^5)$~${\rm rad~m^{-2}}$ with a 10\% variation over the course of seven months \cite{michilli_etal_2018}. RM synthesis clearly indicated a single component of polarized emission behind a single extreme RM screen, implying an environment similar to a massive black hole or a supernova remnant. Meanwhile, FRBs showing small RMs of $\mathcal{O}(10)$~${\rm rad~m^{-2}}$ have proven powerful to constrain the IGMF to less than tens of nG \cite{ravi_etal_2016,bannister_etal_2019} as described in \S\,\ref{sec:cosmicweb}. In the SKA era, the study of polarized emission from FRBs is expected to deliver further interesting information. For example, temporal variation of the polarization angle may provide an indication of the rotation of the repeating sources, while the degree of depolarization may constrain the proportions of regular and turbulent magnetic fields along the line of sight.
The RM variations of repeating FRBs on different time scales can indicate the plasma transition near the source or in the intervening medium.

Extreme RMs of $\mathcal{O}(10^5\,\mathrm{rad\,m^{-2}})$ are likely to be found through comprehensive polarization surveys, including in the environments of high-power radio sources. More generally, the detection of sources with extreme RM values can be an exciting pathway to the discovery of exotic sources, and should be taken into account when planning to optimise the SKA's capability for exploring the unknown Universe (e.g., \cite{wilkinson_2015}).

\subsection{Individual Galaxies}

\subsubsection{Normal and Star-forming Galaxies}

Magnetic fields in the ISM of galaxies have an important moderating influence on the star formation process \cite{krumholz_federrath_2019}, and their properties and evolution are therefore integral to the formation and evolution of galaxies. A fundamental question is the origin of the large-scale coherent magnetic fields that are commonly observed in galaxies in the local Universe \cite{beck_2015,han_2017}. These magnetic fields are understood to have been amplified over poorly-constrained timescales from weak seed fields present at an early phase of galaxy formation \cite{arshakian_etal_2009,beck_etal_2012,martin-alvarez_etal_2018,rodrigues_etal_2019}. To probe this amplification history observationally, efforts focus on studying the magnetic fields in high-redshift galaxies \cite{bernet_etal_2008,farnes_etal_2014,mao_etal_2017,sur_etal_2018}, and testing the galactic dynamo theory in the local Universe \cite{beck_etal_2019}. In the local galaxy population, it is crucial to observe objects with a wide range of properties, including differentially- and solid-body rotating disks. The redshift evolution of the strength and order of galactic magnetic fields is a critical aspect to better understand the physics, as is the dependence on key properties such as star formation rate, galactic rotation, and environment. 

Broadband radio observations have proven to be very effective for probing the structure of magnetic fields in galaxies in the local Universe (e.g., \cite{heald_etal_2009,irwin_etal_2012}). Detailed studies of individual galaxies across a broad range of radio frequencies are crucial to probe the interaction between star formation and magnetic field properties (e.g., \cite{mulcahy_etal_2017,mulcahy_etal_2018,irwin_etal_2019,heesen_etal_2019}). When applied to large and well-defined samples, these observations are now probing the typical magnetic structure in galaxy halos, far from the regions of active star formation \cite{krause_etal_2018}, and thereby constraining the processes such as cosmic ray propagation that support and drive the structures at the outskirts of galaxies \cite{heesen_etal_2018}. Current dynamo theory cannot explain certain aspects of galactic magnetic fields, for example in spiral arms and halos \cite{beck_etal_2019}. A comprehensive understanding of the structure and evolution of galactic magnetic fields can improve models of starburst-driven outflows, IGM feedback \cite{bertone_etal_2006,pakmor_etal_2019}, and cosmological structure formation simulations \cite{berlok_pfrommer_2019,vogelsberger_etal_2020}. Magnetic fields in the outermost parts of galaxies may be understood through a dense RM Grid (e.g., \cite{mulcahy_etal_2014,neld_etal_2018}) as will be delivered by the SKA.

Groups and pairs of galaxies are crucial for understanding the unique influence of environments that are denser than for galaxies in the field, but sparser than in clusters. For example, LOFAR's survey capability is facilitating studies of the typical properties in this regime (e.g., \cite{blazej_etal_2019}). Intergalactic magnetic fields have been recently found in nearby galaxy pairs \citep{kaczmarek_etal_2017,basu_etal_2017}. Using the SKA we will be able to probe the magnetic fields in similar systems to uncover the influence of magnetism in small galaxy groups.

\subsubsection{Active Galactic Nuclei}

The scientific investigation of active galactic nuclei (AGN) requires an understanding of the physics of accretion onto the supermassive black hole (SMBH), and in some cases, the launching of powerful relativistic jets that can extend beyond the host galaxy environment and deposit energy and magnetic field into the surrounding intergalactic medium (IGM). It is expected that mass accretion onto the SMBH makes the AGN a time-variable source \cite{ulrich_etal_1997}, while the magnetic field in the disk enables the accretion of material through the magneto-rotational instability (MRI) \cite{balbus_hawley_1991}. Therefore, while it is difficult to place observational constraints on the magnetic field in the accretion disk, time variability of AGN gives a clue to its properties.

Recently, the Event Horizon Telescope (EHT) has confirmed the general understanding of the standard theory of gravity and the mechanism of radiation transfer in the presence of a SMBH \cite{eht1_2019,eht6_2019}. The next step is to detect the emission from the jet base, where the magnetic field is thought to play a key role in the formation of the relativistic jet. In this regard, polarization and RM observations at relatively high frequencies (allowing detection of the high expected RMs) are important to constrain the magnetic field geometry at the jet launching site. On much larger scales, polarization and RM observations have provided support for the theoretically-expected helical magnetic field geometry \cite{gabuzda_etal_2018}.

However, further downstream there may be a transition to a less ordered field geometry, which radio observations can uniquely probe, before the jet terminates (or is disrupted) to generate the classical double-lobed radio galaxy structure. The radio-lobe morphology can be classified into roughly two types, Fanaroff-Riley (FR) I and FR II. It is unknown to what extent the magnetic field of jets or of the surrounding medium affects the morphology of radio lobes. Investigating links between the lobe morphology and the state of the accretion disk can provide new discovery potential. Finally, AGN jets supply a significant amount of energy and magnetic field to the surrounding IGM \cite{furlanetto_loeb_2001,calzadilla_etal_2019,giacintucci_etal_2020} and are thus crucial for a complete understanding of the formation and evolution of the cosmic structure on large scales.

\subsection{Galactic and Sub Galactic Scale Fields}

\subsubsection{The Milky Way}

The Milky Way is a unique object in which to understand magnetic field structures in spiral arms \cite{han_etal_2018}, the Galactic Center \cite{roy_etal_2008} and the Galactic halo \cite{xu_han_2019,sobey_etal_2019} in exquisite detail. Magnetic fields affect Galactic large-scale (kpc) structures through the Parker instability \cite{rodrigues_etal_2016} and the MRI \cite{pakmor_springel_2013} and alter Galactic small-scale (pc) structures through MHD turbulence. The interplay of these large-scale and small-scale magnetic field components provides unique insights into the dynamo mechanism that enhances and maintains magnetism in galaxies \cite{chamandy_etal_2018}. Furthermore, measurements of magnetic helicity can help to constrain dynamo theory \cite{brandenburg_brueggen_2020}. Although the turbulent magnetic field component is expected to be ubiquitous in galaxies \cite{beck_2015}, it is only directly measurable in detail in the Milky Way (see e.g. \cite{gaensler_etal_2011}), possibly also through interpretation of RM spectra (but see \cite{basu_etal_2019b}). Understanding turbulent interstellar magnetic fields will allow us to gauge their influence on star formation, gas heating and dynamics, cosmic ray propagation and other physical processes in the interstellar medium. 

Interstellar objects such as supernova remnants, planetary nebulae, {\sc H\,ii} regions, globular clusters and gigantic radio loops are all shaped by the influence of the Galactic magnetic field (e.g., \cite{costa_etal_2016,west_etal_2017,costa_spangler_2018,abbate_etal_2020}); see also \S\,\ref{sec:ism}. The magnetic content of High Velocity Clouds (HVCs) such as the Smith Cloud \cite{betti_etal_2019} is important for their longevity. Characterizing magnetic fields in all of these objects yields insight into their evolution and impact on their Galactic environment. For the largest of these Galactic structures, single-dish data will be required to supplement the SKA. New all-sky surveys of the diffuse polarized emission from the Milky Way with single dish telescopes have been conducted recently (e.g., \cite{wolleben_etal_2019,carretti_etal_2019}). These surveys provide polarization observations at multiple frequency channels over a broad bandwidth, which allows us to utilize RM synthesis to study the properties of the Galactic magnetoionic medium \cite{dickey_etal_2019,thomson_etal_2019}.

The Galactic magnetic field deflects ultra-high energy cosmic rays thus concealing their origins \cite{farrar_etal_2013,boulanger_etal_2018}, and the polarization signatures that it causes provide significant foregrounds for extragalactic studies such as CMB polarization (e.g., \cite{poidevin_etal_2018,krachmalnicoff_etal_2018,basu_etal_2019,delahoz_etal_2020}) and the Epoch of Reionization. Therefore, detailed knowledge of the magnetic field strength and structure of the Milky Way is essential for a number of extragalactic science drivers. An SKA1-MID RM Grid will provide a detailed view of Galactic magnetic field structure (e.g., \cite{hutschenreuter_ensslin_2020}), including field reversals, the halo field (e.g., \cite{xu_han_2019,sobey_etal_2019}), and other constraints on Galactic magnetic field models, as well as magnetic fields associated with individual structures like bubbles and nebulae such as SNRs, pulsar wind nebulae, {\sc H\,ii} regions, and planetary nebulae. SKA1-LOW will complement SKA1-MID through facilitating broad-band polarimetry and high-precision RM measurements towards diffuse emission and objects in the Milky Way and beyond. The combination of an RM Grid and polarized diffuse emission has been demonstrated to be powerful in understanding extended structures \cite{sun_etal_2015b}, and will in the future be able to reveal small-scale properties of MHD turbulence, as well as a tomographic three-dimensional view of the magnetized ISM, with an even denser RM Grid using the SKA.

\subsubsection{The Interstellar Medium and Star Formation}\label{sec:ism}

Elucidating the role of magnetic fields in the formation and evolution of molecular clouds, and in particular in regulating star formation within those clouds, are key science goals of astronomy.
In the last decades, it has been recognized that magnetic fields play a crucial role in the mechanism of star formation through various astrophysical processes (e.g., \cite{crutcher_2012,li_etal_2014,hennebelle_inutsuka_2019,federrath_klessen_2012,federrath_2015,krumholz_federrath_2019,pudritz_ray_2019,wurster_li_2018,soler_2019}). For example, magnetic fields are linked to the energy dissipation of the ISM allowing molecular clouds to form, and determining the physical condition of MHD turbulence \cite{brandenburg_subramanian_2005,houde_etal_2011,burkhart_etal_2012,falgarone_etal_2015,federrath_2016}.
A full understanding of MHD turbulence is required in order to explain the formation of dense molecular clouds.
The SKA will be able to probe magnetic fields in molecular clouds and starless cores \cite{padovani_galli_2018}. Magnetic fields provide an additional force to the ISM as a tension, and an additional heating term though ambipolar diffusion. These are key factors required to understand the collapse of a dense molecular cloud to form a protostar.

The evolution of molecular clouds is partially controlled by magnetic fields, including in dense filaments which are the main sites of star formation \cite{andre_etal_2014}. The role of magnetic fields in forming and shaping these is not yet clear \cite{soler_2019} but can be resolved by determining the strength and degree of order of magnetic fields, and comparing with simulations.
Within molecular clouds field strength measurements are via the Zeeman effect \cite{crutcher_kemball_2019}, readily observable in maser emission, but requiring very deep, pointed observations with high spectral resolution at radio frequencies for thermal emission (e.g. OH at 1665 \& 1667 MHz). The Zeeman effect can be prominent in paramagnetic molecules, particularly OH, but the nature of the emission can make measurements difficult, particularly thermal emission for which it is small relative to the linewidth, and thus difficult to measure. Magnetic field structure within molecular clouds is thus typically measured via thermal dust emission, and has been mainly an area for sub-millimetre and far-infrared observations (see, e.g., \cite{poidevin_etal_2020}), but maser emission studies can be productive (e.g., \cite{green_etal_2012,ogbodo_etal_2020}).
RM Grids can also be used to determine the magnetic field in molecular cloud regions, if they are sampled by a sufficient density of sources (e.g., \cite{tahani_etal_2018,tahani_etal_2019}).

Magnetic fields collimate the molecular jets of young stellar objects (YSOs) through magnetic tension \cite{shang_etal_2006,gerrard_etal_2019}. Magnetic fields are also believed to play a significant role in the formation and evolution of planet-forming disks \cite{hu_etal_2019,mori_etal_2019}. The SKA will test this through sensitivity to polarized emission from large grains aligned with the magnetic field, which will also probe grain growth in these disks \cite{hull_etal_2018,guillet_etal_2020}. Magnetic fields in the ISM surrounding star-forming regions both influence the evolution of, and may also be impacted by, {\sc H\,ii} regions \cite{stil_etal_2009,costa_spangler_2018}. The SKA will be able to probe the detailed physical conditions in and around {\sc H\,ii} regions \cite{padovani_etal_2019,meng_etal_2019}. Supernova shock waves appear to significantly amplify magnetic fields of the upstream ISM. The shapes, polarization, and Faraday depth characteristics of SNR can reveal the magnetic fields in the remnant and in the surrounding ISM \cite{kothes_brown_2009,west_etal_2016}. Understanding the magnetic field amplification mechanism is necessary to examine the role of supernova shock waves in the origin of Galactic cosmic-rays.

Additionally the near-Earth environment can be traced through Faraday rotation probes of coronal mass ejections (CMEs; e.g., \cite{howard_etal_2016,kooi_etal_2017}), and studies of the ionosphere through the resulting distortions in source positions (e.g., \cite{hurley-walker_hancock_2018}) and eventually RMs (e.g., \cite{sotomayor_etal_2013,porayko_etal_2019}).

\section{Pathfinder progress}\label{sec:pathfinders}

SKA pathfinders and precursors feature new technical capabilities, as well as design enhancements in comparison with traditional radio telescopes. Many of these facilities are planning substantial new magnetism surveys, which provides us with a twofold opportunity: first, to test many of the analysis techniques that are planned for SKA data products but on a somewhat smaller scale; and secondly to further develop the science questions that will be addressed in the SKA era. In this section we aim to highlight some SKA precursor and pathfinder telescopes that are driving substantial progress in the field, and to capture recent results that they have produced.

\subsection{Science capability from new technology}

Substantial developments with three classes of radio telescope technologies have driven recent improvements in science capability. Broadly speaking, these new telescopes provide increases in sensitivity, field of view, accessible bandwidth, and often combinations of these.

\subsubsection{Phased Array Feeds}

Phased Array Feeds (PAFs) are integrated collections of dual-polarization dipoles, mounted in the focal plane of a dish reflector, with the aim to completely sample the electric field intercepted by the antenna system. Beamforming techniques are used to form `digital' beams on the sky, and simultaneous beams can be used to instantaneously image a far larger field than would be otherwise possible through the use of a single receiver on the same reflector. Two prominent examples of this approach are the Aperture Tile In Focus (Apertif) upgrade to the Westerbork Synthesis Radio Telescope (WSRT) in the Netherlands \cite{oosterloo_etal_2010}, and the Mk-II PAFs on the Australian Square Kilometre Array Pathfinder (ASKAP) \cite{johnston_etal_2008} at the Murchison Radio-astronomy Observatory (MRO) in outback Western Australia.

These PAF-based systems deliver very large fields of view: for example, 9.5 and 30 square degrees with Apertif and ASKAP, respectively, with the difference between them largely driven by the antenna diameter (25 and 12~m respectively). Linking the individual antennas as an interferometric array ensures that the increase in field of view is achieved while retaining excellent angular resolution (typically, $\approx\,5-30^{\prime\prime}$). Additionally, the digital PAF beamforming process is in principle highly flexible and offers the prospect for constraining the orthogonal components of polarization (these systems work on the basis of linear polarization, so `X' and `Y') to have very similar beam shapes, and thereby providing extremely high polarimetric performance throughout the wide field of view.

PAFs are powerful for new magnetism surveys primarily because of their large field of view, which translates to a very high survey speed despite their typically lower sensitivity as compared to the new single-pixel receivers (described in \S\,\ref{sec:wbspfs}). PAF systems also tend to deliver moderately large fractional bandwidth, with practical limits set by the desire to mosaic the same field of view with beams with frequency-dependent sizes, as well as realistic data management and processing constraints.

An example of the polarization performance of ASKAP is the mapping of the southern lobe of Centaurus A \cite{anderson_etal_2018}. The excellent angular resolution, surface brightness sensitivity, and polarization performance together provided a new and highly detailed view of the magnetoionic structure associated with this iconic radio galaxy.

\subsubsection{Wideband single pixel feeds}\label{sec:wbspfs}

Modern single pixel receivers (here referred to as Wide-Band Single Pixel Feeds or WBSPFs) are sensitive feeds across a very broad bandwidth. The newest of these receivers provide a performance comparable to the best traditional `octave' feeds (typical ratios of 1.85:1), but consistently across much broader bandwidths (ratios of 3:1 or greater). Prominent examples of the octave variety and relevant for new polarization surveys are MeerKAT, currently operating at L-band with $800\,\mathrm{MHz}$ bandwidth ($880-1680\,\mathrm{MHz}$)\cite{jonas_etal_2016,camilo_2018} and 
soon to be equipped with systems from $580-3700\,\mathrm{MHz}$; and the Karl G. Jansky Very Large Array (VLA) which has continuous frequency coverage delivered by several contiguous octave feeds between 1 and 50\,GHz \footnote{\url{https://science.nrao.edu/facilities/vla}} \cite{lacy_etal_2020}.
The Parkes radio telescope, as an SKA Technology pathfinder, currently has an ultra-wide bandwidth receiver operating from 700~MHz to 4~GHz \cite{hobbs_etal_2020}. Parkes is also designing a comparable pair of systems to operate from 4 to 26~GHz.

A recent outcome demonstrating the value of the excellent spectropolarimetry that is now available with the VLA is the recognition of compressed upstream magnetised ISM associated with the Sagittarius spiral arm \cite{shanahan_etal_2019} through the detection of polarized extragalactic sources with RMs in excess of $10^3\,\mathrm{rad\,m^{-2}}$. The ultra-wide band receiver at Parkes is already delivering broad bandwidth polarisation studies of pulsars (e.g. \cite{zhang_etal_2019,dai_etal_2020}) and there are also a number of ongoing projects focused on broadband continuum polarisation mapping.

\subsubsection{Aperture Arrays}

Aperture arrays have seen a dramatic resurgence in recent years. These are commonly used at low frequencies ($\nu < 350\,\mathrm{MHz}$). Aperture arrays are similar to PAFs, but are fixed on the ground rather than illuminated by a reflector. Key examples are the LOw Frequency ARray (LOFAR) \cite{vanhaarlem_etal_2013} and the Murchison Widefield Array (MWA) \cite{tingay_etal_2013,wayth_etal_2018}.

Like PAFs, some aperture arrays can be used to observe multiple simultaneous fields of view across very wide areas, leading to flexible survey capabilities. Regardless of individual details, aperture arrays provide very wide fields of view and thus high survey speeds. They are very powerful for magnetism surveys because they provide extremely large fractional bandwidth, at a frequency range distinct from traditional work in the GHz regime.

Aperture arrays have proven to be productive for studying magnetism in the local and distant Universe, both with the MWA \cite{lenc_etal_2016,lenc_etal_2017,riseley_etal_2018} and LOFAR \cite{neld_etal_2018,vaneck_etal_2018,sobey_etal_2019,osullivan_etal_2020}. 
Although radio sources tend to be substantially depolarized at low frequencies in dense environments, the detection rate is better in outskirts of galaxy clusters or for field galaxies, and the precision with which Faraday rotation measures can be determined is excellent. Results from large-area extragalactic polarization surveys with both the MWA and LOFAR are discussed in \S\,\ref{sec:projects}.

\subsection{Upcoming magnetism projects}\label{sec:projects}

\begin{figure}
    \centering
    \includegraphics[width=\textwidth]{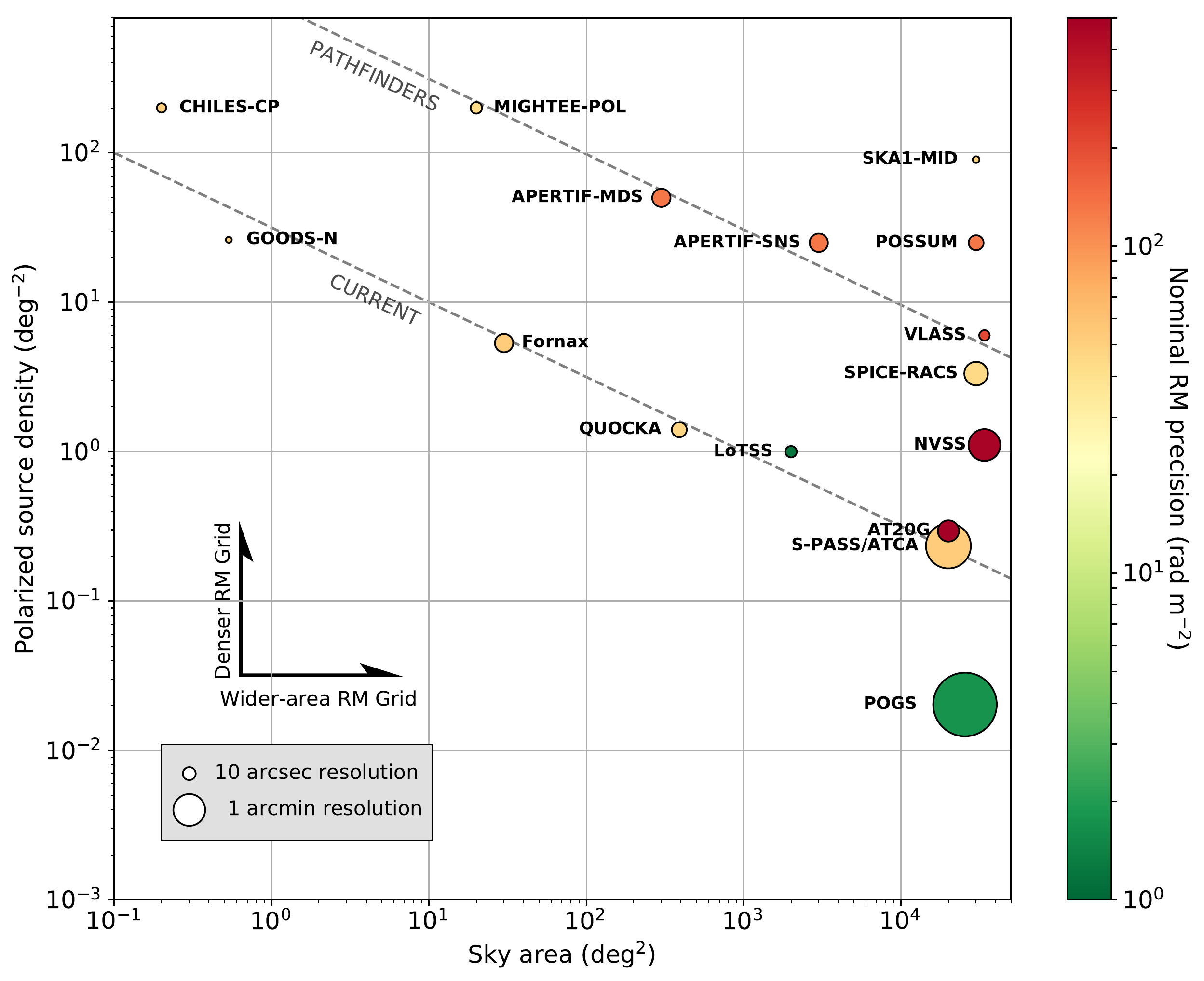}
    \caption{Illustration of the RM Grid survey strengths of various existing and future radio surveys. The size of each marker reflects the angular resolution of the survey, and the colour indicates the nominal RM precision that can be reached, where green indicates the capacity for measurements with lower RM uncertainties. Diagonal dashed lines are intended to highlight the locus of ``current'' (lower) and ``pathfinder'' (upper) survey capability.}
    \label{fig:survey_plot}
\end{figure}

In Figure~\ref{fig:survey_plot}, we demonstrate the continual and ongoing improvement of current and future surveys for the production of RM Grids. Each survey is represented by an indicative sky area and corresponding polarized source density. Additionally, the size of each marker reflects the angular resolution of the survey, and the colour indicates the nominal RM precision that can be reached. Diagonal dashed lines are intended to highlight the locus of ``current'' (lower) and ``pathfinder'' (upper) survey capability. The SKA1-MID survey stands out by having an exceptionally high source density over the full Southern sky, together with excellent angular resolution and RM precision. We now provide brief descriptions of the individual surveys.

{\bf ASKAP RACS.} The Rapid ASKAP Continuum Survey (RACS; McConnell et al., in prep) is the first all-Southern sky imaging survey completed using the full ASKAP array of 36 dishes. The primary goals of RACS are to establish an initial broadband sky model for calibration purposes, and to provide a state-of-the-art continuum survey of the Southern sky. RACS is assembled from  shallow, 15\,min pointings, covering declinations $-90^\circ<\delta<+50^\circ$ with a target angular resolution of $12^{\prime\prime}$. Despite the short integrations, the RMS noise in Stokes $I$ is $\sim300\,\mu\mathrm{Jy\,beam^{-1}}$, a significant improvement over previous surveys such as NVSS \cite{condon_etal_1998} and SUMSS \cite{bock_etal_1999}. A polarization component is being delivered in addition to the total intensity images and catalog. This is called Spectra and Polarization In Cutouts of Extragalactic Sources from RACS (SPICE-RACS; Thomson et al., in prep). It will deliver RMs for a projected 100,000 radio sources, initially in the frequency range $744-1032\,\mathrm{MHz}$, corresponding to a Faraday resolution of $44\,\mathrm{rad\,m}^{-2}$, at an angular resolution of $25^{\prime\prime}$. This frequency range will later be extended to include the $\sim 1150-1450\,\mathrm{MHz}$ range, allowing for greater sensitivity to Faraday thick structures.

{\bf ASKAP POSSUM.} The Polarization Sky Survey of the Universe's Magnetism (POSSUM)\footnote{\url{http://askap.org/possum}} \cite{gaensler_etal_2010} is ASKAP's full-depth polarization survey. It is complementary to the Evolutionary Map of the Universe (EMU) survey \cite{norris_etal_2011}. POSSUM has long planned to provide frequency coverage from about $1130-1430\,\mathrm{MHz}$, but may cover lower frequencies (similar to the initial RACS band as described above), depending on survey commensality and RFI avoidance. POSSUM will generate an RM Grid over a sky area of 30,000 square degrees, through imaging to a depth of about $20\,\mu\mathrm{Jy\,beam}^{-1}$, resulting in up to about a million polarized sources.
The angular resolution will be $\approx10-15^{\prime\prime}$, and depending on final frequency coverage will deliver RM precision better than $4-13\,\mathrm{rad\,m^{-2}}$ for sources detected with signal-to-noise ratios in polarization of at least 5. POSSUM will provide excellent surface brightness sensitivity to map extended emission in, for example, cluster relics and the lobes of radio galaxies.

{\bf ATCA QUOCKA.} The QUOCKA (QU Observations at Cm wavelengths with Km baselines using ATCA) Survey\footnote{\url{https://research.csiro.au/quocka/}} (Heald et al., in prep) will supplement our understanding of the polarized sources that will be detected with POSSUM, through targeted broadband observations of radio galaxies already detected in polarization during the ASKAP early science period. Using the ATCA, $\approx 550$ sources were observed from $1.1-3.1$ and $4.5-8.5\,\mathrm{GHz}$ in several snapshots. These data will be used to develop a clear picture of the (de)polarization properties of radio galaxies and to inform the relatively narrower-band data that ASKAP will deliver. Because of the excellent polarization characteristics of ATCA, QUOCKA will also deliver a high-quality search for circular polarization probing the jet properties within the target sources.

{\bf VLASS.} The VLA Sky Survey (VLASS)\footnote{\url{https://science.nrao.edu/science/surveys/vlass}} \cite{lacy_etal_2020} is an effort to deliver a high-quality catalog along with maps of radio sources over a bandwidth of $2-4\,\mathrm{GHz}$. VLASS is conducted in an on-the-fly survey mode and includes a polarization component as a high priority goal. VLASS has some science aims similar to QUOCKA, and adds well-resolved polarization images and a search for extreme RM sources, up to $\approx 16,000\,\mathrm{rad\,m}^{-2}$. In addition to RMs for compact sources, the high angular resolution ($\approx 2.5^{\prime\prime}$ for total intensity, $\approx 5^{\prime\prime}$ for polarization) together with the broad bandwidth will allow a detailed study of the magnetoionic structures in over 10,000 radio galaxies across the Universe. Figures 12 and 13 in \cite{lacy_etal_2020} show the power of this high resolution for radio galaxy polarization studies. 

{\bf MeerKAT.} Several MeerKAT Large Survey Projects (LSPs) incorporate a polarization component, including in imaging mode: the MeerKAT International GHz Tiered Extragalactic Exploration (MIGHTEE) \cite{jarvis_etal_2016}; commensal polarimetry with the deep {\sc H\,i} LADUMA (Looking at the Distant Universe with the MeerKAT Array)\footnote{\url{http://www.laduma.uct.ac.za/}} project \cite{Blyth_etal_2016}; and MHONGOOSE (MeerKAT {\sc H\,i} Observations of Nearby Galactic Objects --- Observing Southern Emitters)\footnote{\url{https://mhongoose.astron.nl/}} \cite{deblok_etal_2016}. 
The MIGHTEE project team is planning complementary observations with the 
upgraded Giant Metrewave Radio Telescope (GMRT) to provide combined deep polarimetry of several square degrees 
with semi-contiguous frequency coverage from $300-2500\,\mathrm{MHz}$.
The deep field surveys will uniquely permit the study of magnetism in the very faint radio source population, expected to be dominated by distant star-forming galaxies. 
MHONGOOSE will probe magnetic fields in nearby galaxies through direct detection of diffuse synchrotron radiation, as well as the analysis of RMs associated with background radio galaxies.
Additionally, the MeerTime project will regularly observe over 1000 radio pulsars using the MeerKAT array for a wide range of science cases, including probing the magneto-ionic ISM (e.g., \cite{bailes_etal_2018, johnston_etal_2020}).
Finally, the MeerKAT S-band Galactic plane survey, a MPIfR---SARAO collaboration, will deliver polarization products tracing the Milky Way's magnetic field.

{\bf CHIME.} The Canadian Hydrogen Intensity Mapping Experiment (CHIME) operates in the $400-800\,\mathrm{MHz}$ band, with angular resolution $20-40^{\prime}$ and a field of view of ${\sim}200^{\circ}$. It was designed for cosmological investigation of the history of the expansion of the Universe. To reach this goal the Milky Way foreground will have to be subtracted to a precision of $10^{-5}$, implying that very precise imaging of total-intensity and polarized Galactic emission will be produced, covering the entire Northern sky. A subsidiary polarization survey with a 15-m telescope at the Dominion Radio Astrophysical Observatory (DRAO) will be performed over the same band. The 15-m survey will add very extended polarized structure, to which CHIME is insensitive, and will provide polarization calibration. The polarization mapping is a joint project of the CHIME Collaboration and GMIMS (\cite{wolleben_etal_2009}; see below). Subsequent to its construction, CHIME was fitted with a second digital backend for investigation of FRBs and pulsars, including measurement of RMs \cite{chime_2018}.

{\bf LOFAR LoTSS.} LOFAR is opening up the low-frequency ($\nu<200\,\mathrm{MHz}$) polarized sky at high angular resolution ($6-10^{\prime\prime}$), and finding that surveys for linearly polarized sources are more productive at these frequencies than had been anticipated some years ago \cite{lenc_etal_2017,vaneck_etal_2018}. Excellent progress is now being made through the ongoing LOFAR Two-metre Sky Survey (LoTSS) \cite{shimwell_etal_2017,shimwell_etal_2019}. Studies of polarization in individual objects \cite{osullivan_etal_2019} and collections of sources \cite{osullivan_etal_2020,stuardi_etal_2020} hold promise to trace the weak magnetic fields thought to thread the large-scale structure of the Universe (see also \cite{vernstrom_etal_2019}). An RM Grid is being produced from LoTSS data, with around one RM measurement per square degree; leading to a full catalog which when complete should contain up to about 10,000 sources. The polarization maps of these sources will be at excellent angular resolution ($6-20^{\prime\prime}$) and will have remarkably good RM resolution ($\mathcal{O}(1)\,\mathrm{rad\,m}^{-2}$). 
Complementary information will be provided by the WEAVE-LOFAR Survey \cite{Smith2016} through observations in the optical band using the William Herschel Telescope Enhanced Area Velocity Explorer (WEAVE) spectrograph. The knowledge of the redshift of polarized radio sources will provide the means to evaluate the pathlength traversed by the signal along the line of sight, and thus allow inference of the magnetic field properties through statistical approaches and detailed studied of individual sources.

{\bf MWA POGS.} In the Southern sky, the MWA is also developing an all-Southern sky view of extragalactic polarized sources. Using data from the GaLactic and Extragalactic All-Sky MWA (GLEAM) Survey \cite{wayth_etal_2015,hurley-walker_etal_2017,hurley-walker_etal_2019}, a catalog of $\sim 500$ sources has been developed (see \cite{riseley_etal_2018}).
This catalog covers the entire sky South of Declination $+30^{\circ}$, in the frequency range $169-231$\,MHz, for a nominal RM resolution of order $2.6\,\mathrm{rad\,m}^{-2}$. The POlarised GLEAM Survey (POGS) provides high-precision RMs for some 484 extragalactic radio sources and 33 known pulsars \cite{riseley_etal_2020}. However, the low frequency and moderate resolution ($\approx3^{\prime}$) of POGS results in strong beam depolarisation due to fluctuations, including in the Galactic foreground, on scales below the PSF. Polarimetry with the Phase II MWA \cite{wayth_etal_2018}, which provides a factor $\sim2$ improvement in resolution \cite{beardsley_etal_2019} will provide a big step forward in our understanding of the low-frequency polarised sky. This will not only come through reduced beam depolarisation and improved sensitivity, leading to an increased number of source detections in this comparatively unexplored sky area, but also through direct comparison of our Phase I and Phase II measurements, which will allow us to probe the scale size of Galactic magnetic field fluctuations.

{\bf GMIMS.} The Global Magneto-Ionic Medium Survey (GMIMS) has set out to map the polarized emission from the entire sky, North and South, covering $300-1800\,\mathrm{MHz}$ with thousands of frequency channels, using large single-antenna telescopes \citep{wolleben_etal_2009}. GMIMS is mapping the sky in Faraday depth, making the first applications of RM synthesis to single-antenna data. The chosen frequency coverage provides a resolution in Faraday depth of 
${\sim}4\thinspace{\rm{rad}}\thinspace{\rm{m}}^{-2}$ and an ability to detect Faraday depth structures up to ${\sim}110\thinspace{\rm{rad}}\thinspace{\rm{m}}^{-2}$ in extent.
For technical reasons the GMIMS band is broken into three bands, $300-800$, $800-1300$, and $1300-1800\,\mathrm{MHz}$; all-sky surveys are being conducted in each band. To cover both hemispheres, the entire project comprises six component surveys. Three are completely observed: $300-900\,\mathrm{MHz}$ and $1300-1800\,\mathrm{MHz}$ with the Parkes 64-m Telescope, and $1270-1750\,\mathrm{MHz}$ with the DRAO 26-m Telescope. To date only the range $300-480\,\mathrm{MHz}$ from Parkes has been published \citep{wolleben_etal_2019} but processing of the other two is well underway. The GMIMS project is designed to probe the magnetic field as a significant energy-carrying component of the ISM \citep{Ferriere_2001,Heiles_Haverkorn_2012}, and to better understand ISM processes by including magnetic fields.

\section{Technical considerations}\label{sec:tech}
New techniques and methods are required to turn the raw data collected by SKA pathfinders and precursors, and eventually the SKA itself, into useful data products from which we can extract brand-new magnetism science results. In this section, we highlight several examples of new polarization-specific techniques and methods, following the order of data flow: from raw visibilities to high fidelity wide-field full-Stokes image cubes; from image cubes to enhanced polarization products; and from enhanced polarization products to magnetism science outcomes. 

\subsection{Calibration and Widefield Imaging}\label{sec:calim}

Most polarization surveys to be conducted with SKA pathfinders will consist of a mosaic of images formed from distinct telescope pointings, or from on-the-fly mapping, to cover the region of interest instead of a targeted on-axis approach. Therefore, to obtain high fidelity full-Stokes image cubes, we need to be able to extract reliable polarization information from sources away from the bore-sight -- thus the full-Stokes primary beam response needs to be taken into account. While on-axis instrumental polarization can be calibrated and corrected for using standard calibration procedures \cite{hales_2017}, off-axis instrumental polarization calibration is more complex \cite{lenc_etal_2017,riseley_etal_2018,eyles_etal_2020}. Without proper off-axis instrumental polarization calibration, sources that are intrinsically unpolarized will artificially appear to be polarized, and sources that are in fact polarized will display polarization properties deviating from their true intrinsic values. This instrumental effect is a major limiting factor for robust RM determination across wide fields with current radio telescopes.

Correcting for these instrumental response terms is not only important for magnetism science, it is also highly relevant for EoR studies since other Stokes parameters can leak into Stokes $I$, and, especially when the leakage signals have spectral dependencies, these can mimic or distort EoR signals \cite{asad_etal_2018}. It will be important for the SKA antenna design to start with good cross polarization behavior, but corrections will nonetheless be needed. The observed Stokes $IQUV$ is the outer product of the direction-dependent M\"uller matrix characterising the primary beam response, and the true Stokes $IQUV$ (e.g., \cite{smirnov_2011}). The correction is less complicated if the primary beam pattern does not rotate with respect to the sky throughout the observation and in that case can potentially be corrected in the image plane. In general, to fully solve the problem requires a full-Stokes M\"uller $A$-projection treatment, where the primary beam response in full Stokes is projected out and corrected for in the imaging step \cite{bhatnagar_etal_2013}. In order to use this algorithm, well-characterized frequency-dependent beam models are required, using EM simulation results, holographic measurements, or a hybrid of these methods. Self-calibration is an alternative approach to correcting for wide-field instrumental polarization effects. For example, \cite{lenc_etal_2017} tested this approach on MWA data: as sources drift across the field of view, assuming they are unpolarized, one could map out a leakage surface and subtract it out from the observations. These techniques allow for high dynamic range, full-Stokes image cubes with minimal instrumental effects across wide fields of view.

Another important polarization-related calibration step is ionospheric RM correction. This is especially relevant at low frequencies (below $\sim1\,\mathrm{GHz}$) where the nominal uncertainty in Faraday rotation is less than the magnitude of ionospheric Faraday rotation (which can often take values up to at least $\sim\,1-2\,\mathrm{rad\,m^{-2}}$ \cite{sotomayor_etal_2013,mevius_etal_2016,lenc_etal_2016} and becomes more significant at solar maximum). In addition, over long integrations at low frequencies, the time-variable ionospheric RM can depolarize astrophysical signals and must therefore be corrected. These corrections are required in order to allow for magnetism projects that aim to detect very small astrophysical RM, or very small variations thereof. This includes: probing the magnetic power spectrum on very small ($<\mathrm{pc}$) scales using pulsar proper motion and the associated RM time variabilities, the heliospheric RM, IGM magnetic fields at high redshifts, as well as the study of ionospheric properties (e.g., \cite{loi_etal_2015}).

Ionospheric RM is usually approximated as the integral of the product of electron content and ionospheric magnetic fields in a thin shell approximation. Recently, the performance and the accuracy of publicly available global ionospheric maps have been rigorously compared using LOFAR pulsar RM observations \cite{porayko_etal_2019}. While predictions using different geomagnetic fields mostly agree with each other, the accuracy of ionospheric RMs is dominated by the ionospheric total electron content maps: the JPLG \cite{mannucci_etal_1998} and UQRG maps \cite{orus_etal_2005} are found to be superior. It is certainly beneficial to have local high-cadence TEC measurements \cite{malins_etal_2018}. Alternatively, it has also shown that one can use brightly polarized sources, or the bright diffuse polarized synchrotron Galactic emission to track the ionospheric RM if one has polarization measurements of the field of interest at different epochs \cite{lenc_etal_2016,brentjens_2018}. 
    
\subsection{Polarization-specific processing}

Once we have the fully-calibrated Stokes $IQUV$ image cubes in hand, the development of a robust RM Grid catalog that can be used to probe the science questions in \S\,\ref{sec:science} demands that we ask the following basic questions: Where are the polarized sources? What are their angular extents? What are their polarization properties and how complex are their polarization behaviors as a function of wavelength ($\lambda$)? The first two questions can be addressed by source finding in polarization, while the latter two questions can be addressed by broadband polarization analysis tools such as RM synthesis \cite{burn_1966,brentjens_debruyn_2005}, Stokes $QU$ fitting \cite{farnsworth_etal_2011,osullivan_etal_2012} and classification of the degree of complexity exhibited in the Faraday dispersion functions (or ``Faraday spectra''). All these tasks have to be accomplished with a reasonable amount of computing time and resources. Source finding in polarization is highly non-trivial \cite{hales_etal_2012a,hales_etal_2012b}: Noise in polarized intensity is non-Gaussian, individual sources can display polarized emission at more than one Faraday depth, and the peak in polarized intensity does not always coincide with that in Stokes $I$.

Performing source finding on Faraday depth cubes from RM synthesis might not be the most computationally effective approach because the sky is largely empty of compact polarized sources. One of the possibly more efficient methods to find polarized sources is the use of so-called Faraday moments \cite{farnes_etal_2018} which are the mean, standard deviation, skewness, and excess kurtosis of the observed Stokes $Q$, $U$ and polarized intensity (PI) cubes. Another possibility is to analyse only subregions near total-intensity sources of interest \cite{neld_etal_2018}. In the former case, moment maps of the image cubes are produced and then normal source-finding software can be employed on these maps to find polarized sources. In both cases, RM synthesis will be performed on small subregions. This procedure greatly reduces the number of pixels for which one needs to perform RM synthesis, and hence decreases the required computing time, data storage and network transport. The Faraday moment approach has been shown to be produce satisfactory completeness and can bypass the need for procedures that deal with non-Gaussian noise in polarized intensity. The more general source-finding problem is being actively tackled by various SKA pathfinder project teams, including ASKAP's POSSUM, which has a dedicated group that works on data challenges to assess the completeness and reliability of source finding strategies.

Further advances have been made in broadband polarization analysis tools in recent years. RM synthesis is a non-parametric approach to developing a model of source polarization properties from the observational data. Several implementations of the technique have been developed over the past fifteen years, mostly based on direct Fourier transforms (DFTs) but in at least one case through gridding the $\lambda^2$ data and performing a Fast Fourier Transform (FFT)\footnote{\url{https://github.com/mrbell/pyrmsynth}}. Improved data quality (reduced errors and artifacts) should come about from combining the two-dimensional Fourier transform(s) used to form channel images from the visibilities, and the one-dimensional transform embodied in RM synthesis, to form a three-dimensional Fourier transform\cite{bell_ensslin_2012}, a technique which holds interesting promise despite high computational demands. Separately, standard RM synthesis has been accelerated through the use of GPUs \cite{sridhar_etal_2018}. 

Following RM synthesis, the resulting Faraday spectra can be deconvolved through a technique ({\sc rmclean}) similar to cleaning synthesis images \cite{heald_etal_2009}. This technique has been improved by refining the models generated by {\sc rmclean} in a maximum likelihood (ML) framework \cite{bell_etal_2013}. Currently ongoing research is seeking to optimise deconvolution in the channelised image plane for sources with low broadband signal-to-noise ratios.

Stokes $QU$ fitting \cite{farnsworth_etal_2011,osullivan_etal_2012}, on the other hand, is a parametric approach to describe broadband polarization data using models of the magnetized medium along the line of sight. Recent developments in this area include the {\sc firestarter} algorithm \cite{schnitzeler_2018}, which takes into account the spectral indices of each of the fitted polarized components, and the use of convolutional neural networks (CNN) to classify Faraday depth spectra, to distinguish simple sightlines which exhibit only one RM component from more complex sightlines \cite{brown_etal_2019}. Interpretation of Faraday spectra in the case of turbulence can be complicated \cite{basu_etal_2019b} and will require additional consideration.

Measuring polarization over broad bandwidths at low frequencies provides higher resolution in Faraday depth (see e.g. Figure~\ref{fig:survey_plot}). Higher resolution in Faraday depth is vital for distinguishing between discrete Faraday screens and components, as well as subtle differences between RMs \cite{osullivan_etal_2020}, implying tremendous potential value from polarization surveys with SKA1-LOW. At the same time, large RM values are becoming increasingly common within our Galaxy and in association with extragalactic sources such as FRBs \cite{michilli_etal_2018}; understanding these extreme RM values can provide significant insight to the nature of the environment at and around the emission regions. However, large RM values at low frequencies undergo bandwidth depolarization. This depolarization will either cause sources with large RM to be undetected or incorrectly characterised in polarization at low frequencies. The amount of channel depolarization varies with channel width in $\lambda^2$ which can vary substantially over broad bandwidth, making it possible to generate signals that appear to show emission at more than one Faraday depth purely due to an instrumental effect. Recent work \cite{pratley_johnston-hollitt_2019} developed the $\delta\lambda^2$-projection to model and correct the channel depolarization at low frequencies, which is analogous to the projection family of algorithms in interferometric imaging. Furthermore, this work makes it clear that many tools from interferometric imaging may be required for processing polarimetric signals over broad bandwidths.

\subsection{Getting to the Science: Analysis Tools}

Once we have enhanced polarization data products at hand, we need analysis tools that will enable us to deliver the actual cosmic magnetism science goals. For example, an information field theory (IFT)-based framework \cite{Ensslin2009} has been used to reconstruct the Galactic RM sky from noisy measurements of discrete RMs towards background (extragalactic) polarized sources. The algorithm takes advantage of the fact that RMs induced by our Galaxy are spatially correlated, while extragalactic RMs and the observing noise should be spatially uncorrelated. Under this framework, the Galactic RM can be reconstructed, along with an uncertainty map. The original reconstruction \cite{oppermann_etal_2012} has been improved in recent years: first, by relying on fewer assumptions \cite{oppermann_etal_2015}; and also by folding in additional information about thermal gas along the line of sight \cite{hutschenreuter_ensslin_2020}. These algorithms permit a statistical discrimination of the Galactic and extragalactic Faraday rotation. A further isolation of the extragalactic contribution can be accomplished with a Bayesian technique \cite{vacca_etal_2016}, particularly useful for enabling extragalactic magnetism studies. The assumption of an extragalactic uncorrelated term holds when dealing with the largest currently available RM catalog (from NVSS data; \cite{taylor_etal_2009}), characterized by a density of one source per square degree. However, this assumption will not be applicable anymore with the catalog obtained with the SKA1-MID RM Grid, which will be $\approx100\times$ denser, because lines of sight to adjacent sources will be near enough to pass through many of the same media, leading to a correlated extragalactic RM contribution. A complicating factor will be that individual resolved radio galaxies will themselves provide multiple samples of RM.

There are also developments on new tools to identify and characterize structure in images, facilitating comparisons between maps of diffuse Galactic polarized emission and other tracers of the ISM. For example, the rolling Hough transform was first used to detect coherent linear features in {\sc H\,i} maps \cite{jelic_etal_2018}, and is now being applied more broadly to seek correlations between the orientation of magnetic structures and different ISM phases \cite{clark_hensley_2019}. Additionally, the polarization gradient method \cite{gaensler_etal_2011} can be used to constrain fundamental parameters of interstellar turbulence. 

The development of the techniques described in this section has been required to keep up with the rapid capability advances coming from the new telescopes and instruments described in \S\,\ref{sec:pathfinders}. As we progress through the pathfinder era and continue to refine these analysis techniques, we are increasingly preparing for the challenges of the SKA era.

\section{Survey specifications}\label{sec:surveys}

A broadband polarisation survey covering a large sky area and with unprecedented sensitivity and resolution, as we expect to be enabled by the SKA, will allow us to address a broad range of scientific questions as reviewed in \S\,\ref{sec:science}. Advancement in several characteristics of polarisation survey specifications is crucial in order to make substantial progress in measuring the polarised emission of cosmic sources and to gain new insight into the magnetic field of the Milky Way, extragalactic objects, and in the Cosmic Web. A substantial improvement in sensitivity will be required in order to detect more sources (raw sensitivity) and low surface brightness features (importance of short baselines). Excellent angular resolution (importance of long baselines) is important to disentangle source components, for example the lobes of distant radio galaxies, and to minimise the effects of beam depolarisation. Finally, a core aspect of an SKA polarisation survey is the frequency range, which establishes the RM range which can be studied, the degree of Faraday complexity that can be reliably recovered, and the associated precision in measured RMs. Observations at lower frequencies deliver better RM precision, but on the other hand, the fraction of depolarized sources increases towards lower frequencies. In this Section we discuss in detail the specifications of an optimal mid-frequency survey for mapping the polarized sky with the SKA \cite{johnston-hollitt_etal_2015}, followed by additional considerations for ancillary survey activities. 

\subsection{An SKA1-MID RM Grid survey}

{\bf Frequency range.} The radio astronomy community has long experience working within L-band (1-2\,GHz), a frequency range that addresses a broad range of scientific questions from observations of radio continuum and {\sc H\,i}, for example. In the context of radio polarimetry, this is an excellent frequency range because it simultaneously provides reasonable RM precision (typically $\lesssim10\,\rm rad \, m^{-2}$ for sources detected with sufficient signal-to-noise), while many sources that depolarise at lower frequencies are still polarised. Therefore, we plan to carry out a primary polarisation survey using SKA1-MID Band 2, from 950 to 1760~MHz \cite{johnston-hollitt_etal_2015}. In comparison to POSSUM (originally $1130-1430\,\mathrm{MHz}$ although subject to change as indicated in \S\,\ref{sec:projects}) this provides a larger span in $\lambda^2$-space by a factor of 2.7, so that SKA will deliver better RM precision by about the same factor. At the same time, broadband spectral structures including extended features in the Faraday spectrum can be recovered, allowing improved characterisation of Faraday thick sources. Based on the standard expressions \cite{brentjens_debruyn_2005}, SKA1-MID Band 2 will provide a nominal RM precision $\Delta\mathrm{RM}\,\lesssim\,5\,\mathrm{rad\,m^{-2}}$ for sources with signal-to-noise ratios $\mathrm{S/N}\geq5$, and partial sensitivity to resolved structures in Faraday space with breadth up to around $108\,\mathrm{rad\,m^{-2}}$.

The rapidly changing RFI situation bears some brief discussion here. While both SKA sites are protected from ground-based interference, SKA1-MID Band 2 is nevertheless expected to be partially affected by aircraft and satellite communication. In particular, the range from $1025-1150\,\mathrm{MHz}$ is populated by aircraft navigation, while $1217-1251\,\mathrm{MHz}$ are used for Global Navigation Satellite System (GNSS) satellites\footnote{See, for example: \url{https://science.nrao.edu/facilities/vla/docs/manuals/obsguide/rfi}}. Although we expect to be able to mitigate the direct impact of this RFI using standard practices, on the other hand we also expect that the sidelobes resulting from RM synthesis will be substantially increased due to frequency gaps generated by RFI flagging: a na\"ive analysis indicates that the innermost RM sidelobes will increase from $\sim\,30\%$ to $\sim\,50\%$, making reliable deconvolution techniques for RM spectra even more crucial.

{\bf Sensitivity.} To meaningfully increase the number of cataloged polarized sources compared to existing and SKA pathfinder surveys requires a sensitivity of $4\,\mu\mathrm{Jy\,beam^{-1}}$. Assuming that a flux density of five times the noise is required to confidently measure the RM of a source, the polarized flux density threshold for sources in the RM Grid catalogue would be approximately $20\,\mu\mathrm{Jy\,beam^{-1}}$, though we anticipate that we may be able to catalog sources up to a factor of two fainter, depending on advances in data processing, source detection and characterisation \cite{pratley_johnston-hollitt_2016}. Based on studies of the faint polarized radio source population \cite{stil_etal_2014,rudnick_owen_2014}, we expect to be able to find $60-90$ polarised sources per square degree. Preliminary results from POSSUM indicate that ASKAP will be able to measure RMs over the full Southern sky with a density of $25-30$ sources per square degree, in agreement with estimates developed in the same way, hence corroborating the expected density of polarised sources for the SKA polarisation survey.
    
Our current estimate for the observing time required per field in order to reach the target sensitivity at our desired resolution is 15~min \cite{braun_etal_2019}. We expect to require approximately 30,000 pointings to cover the observable sky, and wish to observe at night; thus, the execution of the survey requires $\sim2.5$~years including overhead (see also \cite{johnston-hollitt_etal_2015}). There are aims for developing a scanning mode for MeerKAT, similar to VLASS, which may reduce significantly the overhead.
    
{\bf Angular resolution.} A crucial parameter for SKA polarimetry surveys is the nominal angular resolution. We aim to achieve improvements both over existing surveys, as well as ongoing SKA pathfinder polarisation surveys. Moreover, we aim to achieve a common angular resolution with total intensity continuum surveys in a comparable frequency band. We therefore aim for $2^{\prime\prime}$ resolution, matching the `legacy' reference survey \#4 that is optimised for cross-identification with optical surveys, as described by \cite{prandoni_seymour_2015} (their Table~1). This is significantly better than the expected angular resolution of POSSUM, around $10^{\prime\prime}$, and 2.5 times better than the significantly shallower VLASS's resolution at 3~GHz. This desired improvement in angular resolution is crucial to isolate distinct RM components across a large fraction of resolved sources in the image plane directly. For example, many double radio sources will be clearly separated into two components at $2^{\prime\prime}$ resolution, and more subtle variations across sources will be ubiquitous. The angular resolution is also essential to obtain higher quality RMs as elements of an RM Grid (see \cite{rudnick_2019} for a more detailed assessment of factors crucial for optimising the quality of the SKA RM Grid).

According to a recent simulation \cite{loi_etal_2019b}, the confusion limit in Stokes $Q$ and $U$ for this survey is $0.4\,\mathrm{nJy\,beam^{-1}}$, well below the expected sensitivity level of $4\,\mu\mathrm{Jy\,beam^{-1}}$. This indicates that, in principle, substantially deeper targeted observations in polarization with the same observing setup could be performed without being limited by confusion noise.
    
{\bf ($u,v$) coverage.} A crucial aspect of the expected improvement in SKA polarization survey quality is the far more complete instantaneous ($u,v$) coverage in SKA1-MID observations as compared to current facilities (expected rms of near-in PSF sidelobes is well below 1\%; \cite{ska_lev0}). Even if difficult to quantify in advance over the full survey area, the larger number of baselines will allow a significantly better image reconstruction for complex extended sources and their polarised emission. Although SKA1-MID will itself provide exquisite sensitivity to emission on angular scales up to $\approx0.5-1^\circ$, as an interferometer it will still be subject to missing short spacings, and we anticipate that supplementary single-dish observations will be required for Milky Way research.
  
{\bf Sky coverage.}
We advocate a survey that covers the entire accessible sky from the SKA1-MID site, i.e. 30,000 square degrees. There are a few primary reasons for this preference.
First, a comprehensive understanding of the entire visible Milky Way, through background-source probes and mapping of individual Galactic sources and diffuse emission at all Galactic latitudes, is essential not only for the study of the Galaxy itself, but also to ensure a high-quality foreground model for interpreting extragalactic objects and supporting complementary research programs (including FRBs and cosmological studies).
Moreover, research projects that rely on large numbers of individual sources are best served with wide-area, rather than deep but narrow-area, survey observations. For magnetism studies in particular, robust statistical analysis differentiating galaxy sub-classifications requires large numbers of sources, and probes of redshift evolution also require very large catalogs for unambiguous results.
Finally, support for transient studies (counterpart identification and multi-messenger followup) requires an all-sky foundation.
On the basis of such an all-sky SKA1-MID Band 2 survey, we expect to catalogue up to about 3 million polarized sources \cite{loi_etal_2019b} to form the SKA1-MID RM Grid.  

{\bf Commensality.} As noted above, the SKA1-MID survey described here has the same specifications as the `legacy' reference continuum survey \#4 described by \cite{prandoni_seymour_2015}. However, the relative priority of the reference continuum surveys currently places more emphasis on narrower, more sensitive survey projects. This calls for a more active discussion about survey commensality between science teams, especially those with plans for large-area SKA1-MID surveys, possibly including {\sc H\,i} mapping.
 
{\bf Data products and analysis.} We aim to provide data cubes with sufficient frequency resolution so that we retain sensitivity to the expected RMs. The native spectral resolution of SKA1-MID is 13.4~kHz, which would allow full-sensitivity recovery of emission at RMs up to a few times $10^5\,\mathrm{rad\,m^{-2}}$, expected in some environments as described in \S\,\ref{sec:frbs}. However, by retaining this frequency resolution over the full bandwidth, individual cubes covering a sky area of only 1 square degree would each require 7~TB of disk storage. Moderate RMs of up to about $10^4\,\mathrm{rad\,m^{-2}}$ (more than sufficient for the vast majority of magnetism science cases) would be retained with a more reasonable 1~MHz spectral resolution, with which each cube would only require about 95~GB. At this frequency resolution, IQUV cubes for all pointings for the full RM Grid survey would amount to approximately 5~PB. This data storage aspect is discussed further in \S\,\ref{sec:src}.

RM synthesis and subsequently $QU$ fitting will be used to identify and characterise the RM spectrum for each detected source. Through active research programs on SKA pathfinder telescopes, the community is currently optimising the procedures for this analysis stage to ensure efficient extraction of reliable source characteristics.

\subsection{Additional considerations}

Beyond the specific requirements for our planned SKA1-MID Band 2 RM Grid survey, we note some additional aspects that bear attention in the light of recent progress and results. These may form the basis of updates to our plans as we approach the SKA era, and/or ancillary ideas for additional SKA Key Science Projects.

Dedicated efforts are now underway with SKA pathfinder telescopes (\S\,\ref{sec:pathfinders}, especially headline surveys such as POSSUM and MIGHTEE). These are developing new knowledge and expertise, and as they push into new parameter space they will allow us to learn how to properly develop and utilise these deeper polarization data products. On the basis of the new knowledge that will be gathered between now and the establishment of specific plans to execute SKA surveys, we should be alert to opportunities to improve the survey strategy and the associated plans for data analysis.

As outlined above, a key aspect of polarization surveys is the total frequency coverage, and in particular the total span in $\lambda^2$-space is crucial for maximising the RM precision that can be expected. Moreover, the evolution of the polarized source population over a wide frequency range, along with broadband depolarization and re-polarization behaviours of individual sources, are also of current interest. In the long term, the magnetism community may find that there is strong interest in a further increase in bandwidth by combining surveys across an even larger contiguous frequency range, and in particular SKA1-MID Band 1 ($350-1050$~MHz) to complement the primary Band 2 survey described here.

As noted above, the confusion level for polarization is expected to be far below the noise level that we aim to achieve for the SKA RM Grid survey. This leaves considerable opportunity for deep imaging surveys probing into the faint polarized source population, and probing the evolution of magnetism to high redshift, for example. A deep survey to complement the all-sky RM Grid survey is likely to be of particular interest.

Finally, we note the exciting new polarization results that are presently emerging from LOFAR and its all-sky imaging survey LoTSS \cite{shimwell_etal_2017,osullivan_etal_2020}.
Recently, exploitation of total intensity and polarimetric LOFAR observations has demonstrated that these frequencies are powerful to address the study of magnetisation in as yet unexplored systems. This advance is due to the capability for revealing low surface brightness radio sources (e.g., \cite{govoni_etal_2019,cantwell_etal_2020}) and dispersion in Faraday rotation down to $1\,\mathrm{rad\,m^{-2}}$ or less, likely associated with low-density and weakly magnetized environments, while being completely blind to dense and highly magnetized environments  (e.g., \cite{osullivan_etal_2020,stuardi_etal_2020}). These results highlight the importance of SKA1-LOW observations in combination with SKA1-MID, for a comprehensive study of cosmic magnetism in different magnetic field strength and thermal gas density regimes.
It is clear that there is tremendous opportunity for SKA1-LOW to continue to probe this opening window to much higher sensitivity and over a broader frequency range than is currently being pursued with LOFAR. Plans for such a survey, which would be highly complementary to the MID RM Grid survey, will be developed based on the experience developed from LOFAR and with a view toward commensality with other SKA science areas. 

\section{Looking forward}\label{sec:future}

\subsection{SKA Data Challenges}

The SKA Organisation has recently commenced a program of ``SKA data challenges'', which aim to familiarise the user community with standard SKA data products, and to help working groups develop and provide input into the associated processing and analysis pipelines \citep{bonaldi_etal_2018}. The first data challenge, issued in Nov~2018 and concluded in Jul~2019, consisted of simulated SKA images at 0.56, 1.4 and 9.2~GHz, but only in total intensity and with no spectral index information. As such, this initial simulated data set is not suitable for exploring the various challenges associated with polarimetry and magnetism science. 

Future SKA data challenges will contain polarisation information, hopefully at successively greater levels of complexity. The first step will be to include a spectral index, polarised fraction, polarisation angle and RM for each total intensity component, which will allow the Magnetism Science Working Group to explore basic polarisation pipelines that employ polarisation source finding, RM synthesis or $QU$ fitting. The community has already begun to undertake simple polarisation source challenges in this vein \citep{sun_etal_2015}, which highlight the difficulties that current algorithms experience even with relatively simple situations. A concerted effort will be needed to improve these approaches for SKA.

An initial round of enhancements might include depolarising effects due to multiple RMs within the synthesized beam, Faraday-thick structures, spectral index effects, non-zero synchrotron optical depths, and bandwidth depolarisation within each spectral channel. Further effects will include spatially extended sources (in which the polarised morphology and number of polarised components do not match their total intensity counterparts), the presence of diffuse polarised foregrounds \cite{lenc_etal_2016,vaneck_etal_2017,vaneck_etal_2019}, polarisation leakage (both on- and off-axis), ionospheric Faraday rotation, and averaging of this leakage due to sky rotation or mosaicing. It is important to appreciate that simulating every conceivable polarisation property of an SKA observation is potentially an even bigger challenge than recovering them, and that it is unlikely that any data challenge will fully capture all aspects of the polarised sky.

\subsection{SKA Regional Centres}\label{sec:src}

Data products for the SKA will be made available to users via a network of SKA Regional Centres (SRCs) \cite{ska_src}, which will provide a range of services including archiving, user support, and custom processing and re-processing. There will be a core set of products offered by all SRCs, but also some unique services that perhaps only a subset of SRCs will provide. The aim is that this will be transparent to SKA users, who need not know in which SRC their data are located, and will be able to access their data regardless of where they are based or affiliated.

Magnetism science with SKA will have unique requirements with regard to the data products produced and hosted by SRCs, centred around two main themes. First, magnetism experiments almost always require information in the Stokes parameters $Q$ and $U$, and often also in $I$ and $V$. There will always be leakage between Stokes parameters (see \S\,\ref{sec:calim}), which will be a complex function of frequency, location on the sky, and offset from the pointing centre. Leakage effects will need to be corrected for in calibration and in post-processing, and residual leakage will need to be appropriately characterised so that the user can set thresholds for what constitutes a valid measurement.

Second, the main data-taking mode for magnetism projects will be spectropolarimetry, which has distinct requirements from both continuum (total intensity) and spectral line observations. For most continuum experiments, the behaviour of intensity as a function of frequency will be captured in the form of Taylor terms, which describe the mean intensity, spectral index, spectral curvature and higher-order terms across the band. This is unsuitable for Faraday rotation and other polarimetry experiments, where the behaviour in Stokes $Q$ and $U$ can be highly complicated and oscillatory as a function of frequency, and which cannot be efficiently described as a small set of Taylor terms. Rather, full Stokes frequency cubes will need to be produced, stored, and analysed, so that the behaviour of $Q$ and $U$ as a function of frequency can be studied according to the user's needs. For most spectral line experiments, image cubes will be produced at very high spectral resolution, often over a relatively narrow bandwidth. For polarimetry with SKA1-MID, the spectral resolution needed will be modest ($\ga1$~MHz), but the total bandwidth will be broad ($\gg100$~MHz). For SKA1-LOW, the spectral resolution needs to be higher but the spectral bandwidth is also narrower. In both cases, the sidelobe pattern, resolution and field of view will vary significantly across the band, removing much of the efficiency or commonality in processing that might be employed for a spectral line cube. Furthermore, the dynamic range for many spectral line experiments is fairly low, meaning that deconvolution per channel is relatively unimportant; this will often not be the case for spectropolarimetry.

Some of the corresponding polarisation products that the SRC network will need to make available include:
\begin{itemize}
\item A catalogue of polarisation and Faraday rotation properties for each detected polarised component, using a standard format\footnote{See \url{https://github.com/Cameron-Van-Eck/RMTable} for a proposed format.};
\item A cross-listing of detected polarised components matched to their component or source counterparts from total intensity;
\item ``Coarse'' image cubes of $IQUV$, covering the entire survey area at modest ($\sim1$~MHz) spectral resolution;
\item Possible ``fine'' image cubes of $IQUV$, covering some subset of the sky at high spectral resolution (up to the maximum spectral resolution that will be available over the full bandwidth, 13.4~kHz \cite{braun_etal_2019});
\item Cubes of Faraday depth, either cut-outs around detected sources or covering the entire sky;
\item Images of the peak Faraday depth and associated polarized intensity.
\end{itemize}
The main polarisation catalogue will need to contain a large number of parameters, including: the component's image coordinates; flux, polarised fraction, and position angle at a fiducial wavelength; peak Faraday depth; and uncertainties on all these parameters. Additionally, the catalog needs to characterise the presence of multiple features in the Faraday depth spectrum. 

Multiple developments are taking place worldwide to establish the first elements of the SRC network. For example, a prototype SRC has been built and is running at the Shanghai Astronomical Observatory, Chinese Academy of Sciences \cite{an_etal_2019}, which is open to the community. Similarly, Australia has established a prototype SRC (AusSRC) which is working with the SKA precursor telescopes ASKAP and MWA, providing support for these teams as they develop methods of dealing with large-scale data. In Canada, the Canadian Initiative for Radio Astronomy Data Analysis (CIRADA)\footnote{\url{http://www.cirada.org}} is producing enhanced and science-ready data products for polarization from POSSUM, VLASS and CHIME, and is also acting as a pilot study for an anticipated Canadian SRC. In South Africa the Inter-University Institute for Data Intensive Astronomy \footnote{\url{http://www.idia.ac.za}}, in partnership with several South African institutions, has established a data intensive research cloud\footnote{\url{http://www.ilifu.ac.za}} prototyping SRC technologies as well as processing and analytics tools for SKA pathfinder Large Survey Programs on MeerKAT and the upgraded GMRT.

\section{Summary}

In this contribution we have revisited the science goals and primary survey plans of the SKA Cosmic Magnetism Science Working Group, originally detailed about five years ago \cite{johnston-hollitt_etal_2015}, in the light of the rapid observational progress that is taking place within the field. It is clearly the case that SKA pathfinder telescopes and associated survey efforts are making substantial headway toward the capability that will be required to make the most of the SKA.

In the next few years we will see the emergence of an all-sky RM Grid at intermediate source density from ASKAP POSSUM and the VLASS, taking us partway toward the transformational capability that will be provided by the SKA RM Grid; and exquisite deep polarization images from MeerKAT MIGHTEE, which will inspire even deeper efforts with the SKA. The magnetism community stands at the threshold of a rich observational opportunity.

\vspace{6pt} 



\authorcontributions{writing---original draft preparation, G.H., S.A.M., V.V., T.A., A.D.-S., B.M.G. and M.H.; writing---review and editing, I.A., A.B[asu]., R.B., M.B., A.B[onafede]., T.L.B., A.B[racco]., E.C., L.F., J.M.G., F.G., J.A.G., JL.H., M.H., C.H., M.J.-H., R.K., T.L., B.N.-W., S.P.O'S., M.P., F.P., L.P., M.R., C.J.R., T.R., L.R., C.S., J.M.S., X.S., S.S., A.R.T., A.T. C.L.V., F.V., and J.L.W.}

\funding{The Dunlap Institute is funded through an endowment established by the David Dunlap family and the University of Toronto. B.M.G. and J.L.W. acknowledge the support of the Natural Sciences and Engineering Research Council of Canada (NSERC) through grant RGPIN-2015-05948, and of the Canada Research Chairs program. I.A. acknowledges support by a Ram\'on y Cajal grant (RYC-2013-14511) of the ``Ministerio de Ciencia, Innovaci\'on, y Universidades (MICIU)'' of Spain, and financial support from MCIU through grant AYA2016-80889-P, and through the ``Center of Excellence Severo Ochoa'' award for the Instituto de Astrof\'isica de Andaluc\'ia-CSIC (SEV-2017-0709). A.B[asu]. acknowledges financial support by the German Federal Ministry of Education and Research (BMBF) under grant 05A17PB1 (Verbundprojekt D-MeerKAT). A.B[racco]. acknowledges the support of the European Union's Horizon 2020 research and innovation program under the Marie Sk\l{}odowska-Curie Grant agreement No. 843008 (MUSICA). M.H. acknowledges funding from the European Research Council (ERC) under the European Union's Horizon 2020 research and innovation programme (grant agreement No 772663). M.P. acknowledges funding from the INAF PRIN-SKA 2017 program 1.05.01.88.04. F.P. acknowledges support from the Spanish Ministerio de Ciencia, Innovaci\'on y Universidades (MICINN) under grant numbers ESP2015-65597-C4-4-R and ESP2017-86852-C4-2-R. M.R. acknowledges `Departments of Excellence 2018-2022' grant awarded by the Italian Ministry of Education, University and Research (\textsc{miur}) L.\ 232/2016; Research grant TAsP (Theoretical Astroparticle Physics) funded \textsc{infn}; Research grant ``Deciphering the high-energy sky via cross correlation'' funded by the agreement ASI-INAFn. 2017-14-H.0; Research grant ``From  Darklight to Dark Matter: understanding the galaxy/matter connection to measure the Universe'' No. 20179P3PKJ funded by \textsc{miur}. C.J.R. acknowledges financial support from the ERC Starting Grant ``DRANOEL'', number 714245. Partial support for L.R. comes from U.S. National Science Foundation grant AST17-14205 to the University of Minnesota. J.M.S. acknowledges the support of the Natural Sciences and Engineering Research Council of Canada (NSERC), 2019-04848. X.S. is supported by the National Key R\&D Program of China (2018YFA0404602). S.S. acknowledges support from the Science and Engineering Research Board (SERB) of the Department of Science and Technology (DST), Govt. of India through research grant ECR/2017/001535. F.V. acknowledges financial support of the H2020 initiative through the ERC Starting Grant MAGCOW (714196).}

\acknowledgments{G.H. thanks Phil Edwards for carefully reading the manuscript and providing useful feedback.}

\conflictsofinterest{The authors declare no conflict of interest.} 

\externalbibliography{yes}
\bibliography{magnetism}

\end{document}